\documentclass{aa}
\usepackage{graphicx}
\usepackage{natbib}
\usepackage{amsmath}
\usepackage{multirow}
\usepackage{amsfonts}
\usepackage{amssymb}
\usepackage{mathptmx}
\bibpunct{(}{)}{;}{a}{}{,} 
\newcommand{\Msun}{\ensuremath{\text{M}_{\odot}}}

\usepackage{array}
\newcolumntype{M}[1]{>{\raggedright}m{#1}}

\begin{document}
\title{Reverse dynamical evolution of Eta Chamaeleontis}
\author{Christophe Becker\inst{1}
  \and Estelle Moraux \inst{1} 
  \and Gaspard Duch\^ene \inst{1,2}
  \and Thomas Maschberger \inst{1}
  \and Warrick Lawson \inst{3} 
}
\institute{UJF-Grenoble 1 / CNRS-INSU, Institut de Plan\'etologie et
  d'Astrophysique de Grenoble (IPAG) UMR 5274, Grenoble, F-38041,
  France
  \and Astronomy Department, University of California Berkeley,
  HFA B-20 3411, Berkeley CA 94720-3411
  \and School of Physical, Environmental and Mathematical Sciences,
  University of New South Wales, Australian Defence Force Academy,
  Canberra ACT 2600, Australia
  }
\date{Draft version December 17, 2012}
\abstract{ In the scope of the star formation process, it is unclear
  how the environment shapes the initial mass function (IMF).
  While observations of open clusters propose a universal picture for
  the IMF from the substellar domain up to a few solar masses,
    the young association $\eta$ Chamaeleontis presents 
  an apparent lack of low mass objects ($m<0.1\ $\Msun). Another unusual
  feature of this cluster is the absence of wide
  binaries with a separation $>50$ AU.
  }
  {We aim to test whether dynamical
  evolution alone can reproduce the peculiar properties of the association under the
  assumption of a universal IMF.}
  {We use a pure N-body code to simulate the dynamical
  evolution of the cluster for 10 Myr, and compare the results with
  observations. A wide range of values for the initial parameters
  are tested (number of systems, typical radius of the density distribution and virial ratio)
  in order to identify the initial state that would most
  likely lead to observations. In this context we
  also investigate the influence of the initial binary population on
  the dynamics and the possibility of having a discontinuous single IMF near the transition to the
  brown dwarf regime. We consider as an extreme case an IMF with no low mass
  systems ($m<0.1\ $\Msun).
  }
  {The initial configurations cover a wide range of initial density,
  from $10^2$ to $10^8$ stars/$\text{pc}^3$, in virialized,
  hot and cold dynamical state. We do
  not find any initial state that would evolve from a universal
  \textit{single} IMF to fit the
  observations. Only when starting with a truncated IMF
  without any very low mass systems \textit{and} no wide binaries, can we
  reproduce the cluster core properties with a success rate of 10\% at best.
  }
  {Pure dynamical evolution alone cannot explain the observed properties of
  $\eta$ Chamaeleontis from universal initial conditions. The lack of brown dwarfs
  and very low mass stars, and the peculiar binary properties (low binary
  fraction and lack of wide binaries), are probably the result of the star formation process
  in this association.}


\maketitle 

\section{Introduction}
As an imprint of the star formation process and governing the
evolution of star populations, the initial mass function (IMF) has
been studied in depth in the solar neighbourhood as well as
in young open clusters. Particular interest has been devoted to the
question of the universality of the IMF: is there a unique mass
distribution resulting from the interplay of physical processes of
star formation, or does it vary with gas density, metallicity
\citep{Marks_et_al_2012} or turbulence? \\ 
Introduced by \citet{Salpeter_1955}, the IMF was first described for
stars in the mass range $0.4$ \Msun\ to $10$ \Msun\ as a power law $ \xi(\log\ m) =
\frac{dn}{d\log\ m} \propto m^{-\Gamma}$, with $\Gamma=1.35$ in
logarithmic scale. This field star power-law index was independently
established by \citet{Kroupa_1993} for $0.5$ \Msun\ to $1$ \Msun\ and extended by
\citet{Massey_2003} to $10$ \Msun.
Focussing on close open clusters, e.g. the Pleiades
\citep{Moraux_et_al_2003, Lodieu_et_al_2007}, IC 4665
\citep{de-Wit_et_al_2006}, $\alpha$ Per
\citep{Barrado-y-Navascues_2002}, or Blanco 1
\citep{Moraux_et_al_2007b}, it was possible to explore the system
(i.e. incorporating
both single objects and unresolved binaries) mass function
in the lower mass regime down to $\simeq$ 0.03 \Msun. Investigations on
the shape of the mass function in various environments show some deviations that
can be explained by uncertainties due to e.g. different sampling, dynamical
evolution, and stellar evolution models, but show no evidence for any
significant variation \citep{Scalo_2005,
 Bastian_Covey_Meyer_2010}.These studies lead
to a universal picture of the {\it system}
IMF down to 0.03 \Msun \citep[see review of][]{Kroupa_et_al_2011} as a non-monotonic
function showing a maximum around 0.25 \Msun\ and a power-law tail at
the high mass end. Many functional forms can be tailored to this IMF,
e.g. segmented power-laws \citep{Kroupa_Tout_Gilmore_1993}, a
log-normal function plus a power-law tail \citep{Chabrier_2003}, or a tapered power law
\citep{De-Marchi_Paresce_Portegies-Zwart_2005, Parravano_et_al_2011, Maschberger_2012}. \\
In this paper, the universality of the IMF is investigated by focussing
on the dynamical evolution of the stellar group $\eta$ Chamaeleontis.
Since its discovery by \citet{Mamajek_Lawson_Feigelson_1999} this
cluster has been the target of many observational studies
\citep[e.g][]{Luhman_2004,Brandeker_et_al_2006,Lyo_et_al_2003}. It is a young
\citep[6-9 Myr,][]{Lawson_Feigelson_2001,Jilinski_Ortega_Reza_2005},
close (d\ $\simeq$\ 94 pc) and compact group of 18 systems (contained in a radius of 0.5 pc). 
Its system mass function was found to be consistent with that of other young open clusters
and the field \citep{Lyo_et_al_2004} in the
mass range 0.15-3.8 \Msun , but with a lack of lower mass
members. This challenges the universal picture of the IMF,
unless the observed present day mass function has already been affected by
dynamical evolution.
Despite deep and wide-field surveys \citep{Luhman_2004,
Song_Zuckerman_Bessell_2004,Lyo_et_al_2006}, no very low mass systems
($m\lesssim0.1$ \Msun \footnote{The lowest mass members have
    an estimated 
    spectral type around M5, leading to masses between 0.08 and 0.16
    \Msun , depending on the adopted evolutionary tracks \citep{Lyo_et_al_2003,
Luhman_Steeghs_2004}})
was found within 2.6 pc from the center. A
recent study by \citet{Murphy_Lawson_Bessell_2010} reported the
discovery of four probable and three possible low mass members ($m<0.3$
\Msun) in the outer region, between 2.6 and 10 pc from the
cluster center. This suggests that the lower mass members might have escaped
from the cluster core due to dynamical encounters and lie at
larger radii than the more massive members. Moreover, the cluster appears to be mass
segregated with all the massive stars ($m>1.5$ \Msun) concentrated in
its very central region, which supports the picture of dynamical evolution.
Among the 18 systems 5 are confirmed binaries and 3 are
possible binaries yielding a binary fraction in the range
[28\%,44\%]. As summarised by \citet{Brandeker_et_al_2006}, none of
these binaries have a projected separation greater than 20 AU, and the
probability for a star to have a companion at separations larger than
30 AU was estimated to be less than 18\%. This is opposed to the
58\% wide binary probability in the TW Hydrae association
\citep{Brandeker_et_al_2003}, despite its
similar age. This deficit of wide binaries in $\eta$
Chamaeleontis may also be explained by their disruption through dynamical
interactions.\\ 
In a previous study \citep{Moraux_Lawson_Clarke_2007}, we
considered whether dynamical interactions could explain the lack of
very low mass systems ($m<0.1$ \Msun) in the cluster core, starting
with a universal IMF. We applied
an inverse time integration method by
sweeping the parameter space for the initial state in order to find those that
best lead, as a result of a pure N-body simulation, to the observed
properties of $\eta$ Cha. This method has been applied
in numerous earlier studies \citep[e.g][]{Kroupa_1995a, Kroupa_Bouvier_2003,Marks_Kroupa_2012} to 
obtain a comprehensive picture of the early dynamical evolution of
star clusters. In our case this was designed as a test
of the universality of the IMF. Assuming a log-normal shape for the
{\it system} IMF \citep{Chabrier_2003} we span a large
range of initial densities. We found that it was possible to reproduce
the observations starting from a very dense configuration ($10^8$
stars/$\text{pc}^3$) with a success rate of 5\%. The simulations, however, did not include any
primordial binaries nor considered the creation of binaries in the detailed
analysis. The gas was removed initially and we assumed that
the cluster was in virial equilibrium. \\
In the present study, we follow the same method in an attempt to
reproduce the observed state of $\eta$ Cha, but we now take into account an
initial binary population and its evolution.  In a first set of
models, we assume a universal log-normal IMF \citep{Chabrier_2005}
before considering a possible discontinuity \citep{Thies_Kroupa_2007}
around the substellar limit, and a truncated IMF with no system below
0.1 \Msun. 
The simulations still start after the gas has been expelled but virial
equilibrium is not required.\\
The outline of this paper is as follows: we first discuss the
statistical significance of the deficit of very low mass systems
($m<0.1$ \Msun) in the cluster
core (Section~\ref{stat}) before describing the numerical scheme
adopted for the simulations, especially the initial conditions and the
parameter grid (Section~\ref{In}). The analysis procedure is introduced
in Section~\ref{Analysis}. Section~\ref{Res} presents
the results obtained when starting with a log-normal IMF. We discuss alternative
initial conditions in Section \ref{Disc} before presenting our conclusions.
\section{Statistical issues}
\label{stat}
With less than 20 systems in the cluster core, the statistical analysis of
$\eta$ Cha has to be done carefully, especially when considering a
standard distribution such as the IMF. In the range from 0.15 to 4 \Msun, the
$\eta$ Cha mass function (MF) was found to be consistent with the IMF
derived for young embedded clusters
\citep{Lyo_et_al_2004,Meyer_et_al_2000} and field stars by comparing
the ratio of stars with mass $m>1\ \Msun$ to stars
with mass $0.1<m<1~\Msun$.
\citet{Lyo_et_al_2004} predicted about 20 members with $m<0.15\
\Msun$\ by comparison with the Trapezium MF. None has been found
within a 2.6 pc radius despite deep and wide searches, indicating a
strong deficit of very low mass systems in $\eta$ Cha.
However, using a Kolmogorov-Smirnov (KS) test, \citet{Luhman_et_al_2009} derived a
probability of $\simeq$10\% that $\eta$ Cha is drawn from the same IMF
as Chameleon I or IC348, not revealing significant differences between
those distributions. One can therefore wonder whether the lack of very
low mass objects (single stars/brown dwarfs or unresolved binaries with $m<0.1\ \Msun$,
hereafter VLMOs)
inside a 2.6 pc radius from the cluster
center represents a significant deviation from the universal MF of open
clusters.
\begin{figure}
  \resizebox{\hsize}{!}{\includegraphics{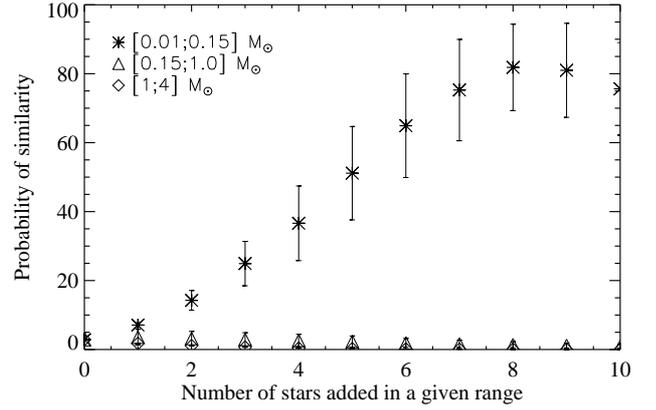}}
  \caption{KS test evaluating the similarity of the $\eta$
    Cha mass distribution with the log-normal system IMF. The comparison is done
    considering three mass ranges, in order to assess which parts
    of the distribution are alike. We progressively added ten stars in
    $\eta$ Cha data set, selected randomly from the IMF within the
    mass range [0.01;0.15] \Msun. Each time a star is added we compute
    a probability from the KS test. This process is repeated for the
    mass ranges [0.15;1] \Msun ~and [1;4] \Msun.}
  \label{Ag_evol}
\end{figure}
If we choose the log-normal fit to the Pleiades system MF
as a reference \citep{Moraux_et_al_2003}, then the KS probability for
testing the hypothesis that the stellar masses in $\eta$ Cha are
chosen from this log-normal MF is 2.8 \%. To assess the sensitivity of
this result to the data set, we present in Fig.~\ref{Ag_evol}
the evolution of the KS probability while systems
are randomly added to the list of known $\eta$ Cha members from three
different mass ranges ([1-4] \Msun, [0.15-1] \Msun\ and [0.01-0.15]
\Msun). The probability increases uniformly to 80\% until eight systems with $m<0.15$ \Msun\
are added, which points out the importance of the deficit of VLMOs
relatively to medium and high mass stars. However the KS probability is already greater than
5\% when only one such system is added, and we cannot reject the
possibility that this data set is drawn from the MF used as
reference. As a result, a paucity of VLMOs may be present
but it might not be statistically inconsistent with the Pleiades MF.\\
Nevertheless, even if the deficit of very low mass systems is not really significant,
it is additional to the other peculiar properties that also need to be understood:
the lack of wide binaries with separation larger than 20 AU, and the presence of mass
segregation.
\section{Numerical set-up}
\label{In}
In this Section we describe the physical properties that we tested in 
our models: the \textit{single}
initial mass function (of all stars counted individually) and the primordial binary properties (binary fraction, separation
and mass ratio distribution). We then review the assumptions
corresponding to the lack of gas treatment and the density profile, and we present
the parameter grid that we used for each model:
the number of systems, the characteristic spatial scale for the
density distribution function, and the global virial ratio.\\
\subsection{IMF}
\label{imf}
As our main hypothesis we choose a \textit{single} IMF
of log-normal form as suggested by \citet{Miller_Scalo_1979}
$$ \xi\text(log\ m)\ \propto\ \exp\left[-\frac{(\text{log}\ m\ -
    \text{log}\ \mu)^2}{2\sigma^2}\right]$$
This function was fitted in the 0.1-1 \Msun ~mass range to the nearby
galactic disk MF based on a volume limited sample within 8 pc and yields $\mu =$ 0.2 \Msun\ and
$\sigma =$ 0.55 \citep{Chabrier_2005}. A similar result was obtained by
\citet{Bochansky_et_al_2010} for the field M-dwarf MF based on a much
larger (several million stars), but unresolved sample.
In order to test its universality, we use the \textit{single} IMF
proposed by \citet{Chabrier_2005} in our models A, B and C. 
Masses were chosen within a mass range
consistent with observation, from 0.01 to 4 \Msun. \\
As an alternative,
we test the possibility that the \textit{single} star IMF may be
discontinuous (model D) with the majority of brown dwarfs following their own
IMF, as suggested by
\citet{Thies_Kroupa_2007}. Since this may result in a lower number
of VLMOs in the cluster initially, we might expect this
initial condition to be more favourable in reproducing the
observations. \\
We also study the extreme case where the IMF is not
universal but truncated in the low mass domain, with no system below 0.1 \Msun ~to follow
the observations (models E and F).
\subsection{Binary properties}
Observations of young clusters reveal a broad range of binary fractions $b_f$,
defined as $\displaystyle b_f\ =\ \frac{N_b}{N_b+N_s}$, where $N_b$ is the number of binaries and
$N_s$ the number of single objects.
According to N-body numerical simulations \citep{Kroupa_1995a}, this
is consistent with a universal initial binary fraction of 100\% that decreases with time
depending on the star cluster density. We thus fixed the initial binary
fraction to $b_f=$100\% for most of our models (except model C and F).\\
Concerning the mass-ratio distribution, we draw two masses
randomly (in Model A) from the same \textit{single} IMF in order to get the primary and
the secondary masses ; the approach adopted
by \citet{Kroupa_1995a}. For all other models (except Model D) we adopted
a flat mass-ratio distribution \citep{Reggiani_Meyer_2011}, and the IMF
used to obtain the mass for the primary had to adjusted (see section \ref{modelB}) so that once
the pairing is done we retrieve the \textit{single} IMF from
\citet{Chabrier_2005} presented above.\\ 
As for the separation distribution, we took the one derived by
\citet{Kroupa_et_al_2011} (see their equation 46) to fit the
observational data for F, G and K field stars
\citep{Duquennoy_Mayor_1991,Raghavan_et_al_2010} using inverse 
dynamical population synthesis and
taking into account early evolution of the orbital parameters for
small period systems. 
The separation distribution ranges from 0.1 AU to $10^5$ AU
for the models A, B, D and E. In
accordance with conclusions of the work by \citet{Kroupa_et_al_2011}, we also chose an
initial eccentricity distribution already in statistical equilibrium
to ensure a thermal final distribution. Although observations
of multiple systems rather support an essentially flat distribution
of eccentricity \citep[e.g][]{Abt_2006}, we note that dynamical simulations
are hardly sensitive to the selected eccentricity distribution
\citep{Kroupa_1995c}. We therefore favour consistency with previous work to
enable direct comparisons.\\
Given the observed lack of wide binaries, we consider the possibility that no
wide binary has formed initially. Instead of assuming a
Kroupa-like separation distribution we also explore a truncated
distribution at large separation together with a smaller binary
fraction (models C and F, see section \ref{SepDist}).\\
A summary of all considered models is given in Table \ref{Models}.
\begin{table}
  \caption{Physical properties corresponding to each model.}
\label{Models}
\centering
\begin{tabular}{M{1.4cm} | c c c}
\hline
& log-normal IMF & discont. IMF & truncated IMF \\
\hline
$b_f=$100\% + random pairing & model A & model D & \\
$b_f=$100\% + flat mass ratio & model B & & model E \\
separation cut-off + flat mass ratio & model C & & model F
\end{tabular}
\end{table}

\subsection{Gas\label{Gas}}
For all our models, we assume that
the gas is already removed at the start of the simulations,
but that the cluster is not necessarily relaxed to virial
equilibrium. We
estimate that our initial state depicts a cluster that is between 0.1 Myr to 3 Myr
old, depending on the picture for gas removal \citep{Tutukov_1978}.
Indeed, there is no
clear consensus about the time scale for gas dispersal, estimated from
0.1 to a few crossing times, depending on the mechanism in
play (OB star wind, supernovae remnant or stellar outflows). In the
case of $\eta$ Cha, the mechanism for gas removal may involve an
external factor on a time scale that could be as long as a few
Myr. \citet{Ortega_et_al_2009} proposed a common formation scenario
for young clusters in the Scorpio-Centaurus OB association ($\eta$
Cha, $\epsilon$ Cha and Upper Sco), by backtracking bulk motions.
In this dynamical picture, $\eta$ Cha was born in a
medium likely being progressively blown out by strong stellar winds
coming out from the Lower and Upper Centaurus Crux complex. Another
possibility to expel the gas from the cluster may involve feedback
from a massive stellar member. \citet{Moraux_Lawson_Clarke_2007}
showed that $\eta$ Cha initial state might have been very compact,
with a crossing time of about $2~10^4$ yrs (for a total mass of 15
\Msun\ and a radius of 0.005 pc). In this extreme case, the
presence of one B8 star (after which the cluster is named) might be sufficient to
remove the gas within $10^5$ years. With a cluster age estimate of
$8\pm1$ Myr, we run the simulations for 10 Myr.
\subsection{Density and velocity distribution}
For all models, the systems are distributed spatially using a Plummer model
$$\rho_{\text{Pl}}(r) = \frac{3 N_{\text{sys}}}{4\pi R_{\text{Pl}}^{3}} \left[
  1 + (r/R_{\text{Pl}})^2 \right]^{-5/2}$$ where $N_\text{sys}$ is the
initial number of systems.\\
The velocities of each individual object are computed according to this density
distribution and to the initial virial ratio $Q_i\ =\
{E_{kin}}/{E_{pot}}$ where $E_{kin}$ is the total kinetic energy of
  the cluster and $E_{pot}$ the gravitational energy. For each model,
  we are thus left
    with three free parameters: the initial number of systems
    $N_{\text{sys}}$, the Plummer radius $R_{\text{Pl}}$ and the virial ratio, $Q_i$.\\
\subsection{Parameter grid}
\begin{table}
  \caption{Parameter grid for model A. We considered all combinations of
    $N_{\text{sys}}$, $R_{\text{Pl}}$, and $Q_i$}
\label{Par}
\centering
\begin{tabular}{c | c c c c c c}
\hline
    $N_{\text{sys}}$ & 20 & 30 & 40 & 50 & 60 & 70 \\
    \hline \hline
    $R_{\text{Pl}}$ (pc) & 0.3 & 0.1 & 0.05 & 0.03 & 0.01 & 0.005\\
    \hline \hline
    $Q_i$ & 0.3 & 0.4 & 0.5 & 0.6 & 0.7 \\
\hline
\end{tabular}
\end{table}
From the shape of the IMF, we estimated an initial value
of $N_{\text{sys}}=50$ by the requirement to have four stars with mass greater
than 1 \Msun. To cover a wide range of densities at fixed radius, we
tested $N_{\text{sys}}$ from 20 to 70. The initial cluster radius was
first estimated to fit a constant surface density derived from
observations of star forming regions \citep{Adams_et_al_2006}, giving
0.3 to 1.0 pc for 50 systems. The study of
\citet{Moraux_Lawson_Clarke_2007} showed that a dense initial configuration
was necessary in order to eject enough members from the cluster core
and reproduce the lack of VLMOs. To favour dynamical
interactions, we took a radius varying from 0.3 to 0.005 pc, yielding
a density range from 500 stars/$\text{pc}^3$ to $10^8$ stars/$\text{pc}^3$. In order to
assess the effect of initial equilibrium, we tested cold,
gravitation-dominated configuration ($Q_i\ =\ 0.3$)
and hot, initially expanding configuration ($Q_i\ =\ 0.7$).\\
For each model (A to F), an initial configuration is characterized by a combination of
\{$N_{\text{sys}}$, $R_{\text{Pl}}$, $Q_i$\} from the values given in
Table~\ref{Par}. In total, 180 arrangements were tested for each model.
\subsection{N-body code}
We use the NBODY3 code \citep{Aarseth_1999} that
performs a direct force summation to compute the dynamical evolution of
the cluster. Close encounters are treated by Kustaanheimo-Stiefel (KS)
regularization for hard binaries \citep{Kustaanheimo_Stiefel_1965},
which uses a space-time transformation to remove the singularity and
then simplify the two body treatment, or chain regularization method
\citep{Mikkola_Aarseth_1990} for few body interactions
(e.g. binary-single star). There is no stellar evolution.
\subsection{Modelling procedure}
The time evolution of the initial conditions described earlier produces
output of positions and velocities for each star
every 0.05 Myr. The NBODY3 output files also provide details for
close binaries (semi-major axis, eccentricity) identified as bound
double systems.The stellar cluster is put in a galactic potential that
defines its tidal radius: $$r_t=\left(\frac{G\
M}{4A(A-B)}\right)^{1/3} \simeq 1.4\ M^{1/3} \ \textrm{pc}$$ where
$M$ is the cluster total mass (in solar masses) and $A$ and $B$
are the Oort constants \citep{King_1962}. Given the parameter grid,
the initial $r_t$ value varies between 3.1
and 4.8 pc (the estimated value for $\eta$ Cha is around 3.5 pc
assuming a total mass of 15 \Msun , \citet{Lyo_et_al_2004}).
Objects are considered as being ejected and then removed from the simulation as soon as
they are further than twice the cluster tidal radius from the
center.\\
For each initial configuration \{$N_{\text{sys}}$, $R_{\text{Pl}}$,
$Q_i$\} we generated 200 simulations, changing only the random seed, for
statistical purposes. Every simulation computed the cluster dynamical
evolution for 10 Myr.\\
\section{Analysis procedure}
\label{Analysis}
In order to retrieve as much information as possible we analyse our
set of simulations in two different ways for each model. First we consider the same
analysis procedure as in \citet{Moraux_Lawson_Clarke_2007} that aims at
finding final states that fit the observational data. Secondly, in
order to better understand the results of the first analysis, we
perform a statistical analysis.
Both methods are based upon a set of constraints
derived from the observations.
\subsection{Observational criteria}
We use a set of criteria described below to evaluate if a simulation
at a given time is close to reproducing the observations. Each criterion
is associated with a range of validity assuming Poisson
statistics: a criterion $i$ is satisfied if $ N_i \in [O_i -
\sqrt{O_i}, O_i + \sqrt{O_i}]$, where $N_i$ is obtained by simulation
and $O_i$ is given by the observations.  A summary of the chosen ranges is
given in Table~\ref{Bla}.
\begin{table}[t!]
\caption{Selection range adopted for the observational criteria}
\label{Bla}
\centering
\begin{tabular}{c c c}
\hline
    Criterion & Range & Restricted to\\
    \hline \hline
    Systems &  $N_1=[14,22]$ & $r<0.5$ pc\\
    \hline
    \multirow{2}{*}{Massive stars } & \multirow{2}{*}{$N_2=[2,4]$} & $r<0.5$ pc\\
      & & $m>1.5$ \Msun\\
    \hline
    \multirow{2}{*}{VLMOs} & \multirow{2}{*}{$N_3=[0,1]$} & $r<2.6$ pc\\
      & & $m<0.1$ \Msun \\
    \hline
    \multirow{2}{*}{Halo} & \multirow{2}{*}{$N_4=[0,1]$} & $0.5<r<10$ pc\\
      & & $m>0.5$ \Msun\\
    \hline
    Binary fraction& $N_5=[22,50]$\%& $r<0.5$ pc\\
    \hline
    \multirow{2}{*}{Wide binaries} & \multirow{2}{*}{$N_6=[0,1]$} & $[50;400]$ AU\\
     & & $r<0.5$ pc\\
    \hline
    Time & [5,8] Myr & - \\
\hline
\end{tabular}
\end{table}
\paragraph{Number of systems ($N_1$)}
To account for the membership and compactness of the core, we consider
the total number of systems in a $0.5$ pc
sphere.
Since 18 systems have been observed within the core radius,
we choose the range of 14 to 22 systems for a simulation to fulfil
this criterion. Unless mentioned otherwise the term \textit{system} refers to a
single object or a binary of any mass within the stellar or substellar domain.\\
To take into account observational limitations in the comparison between simulations and observations, we identify binaries as
closest neighbour pairs in projection (i.e. not necessarily bound)
with a separation smaller than 400 AU (which corresponds
to 4'' at the cluster's distance). At larger separations binaries are
observationally identified as two single objects \citep{Kohler_Petr-Gotzens_2002}.
\paragraph{Number of massive stars ($N_2$)}
Since three systems were found in the central region with a mass $m > 1.5$
\Msun, we require to have between $2$ and $4$ of them in the
simulations. When counting massive stars, a binary system is
considered as a single object with a mass corresponding to its total mass.
\paragraph{Number of systems in the halo ($N_3$)}
No potential cluster member has been identified by the ROSAT All Sky
Survey (sensitive to late-K type stars) outside the cluster core up to
a distance of 10 pc. This translates into the following criterion:
less than one cluster member more massive than 0.5 \Msun\ must lie within
the distance range [0.5-10] pc from the cluster center.\\
Recently \citet{Murphy_Lawson_Bessell_2010} have discovered four probable
and three possible less massive members (in the spectral range K7 to
M4, i.e. $0.1 \Msun~<m<0.3$ \Msun) at a distance between 2.6 pc and 10 pc from
the cluster centre. However, since the
status of these candidates is not confirmed, we will check a posteriori
that some simulations matching all other criteria do produce a number
of low-mass halo stars that is consistent with the small number suggested
by Murphy's study.
\paragraph{Number of very low mass objects ($N_4$)}
No system with $m< 0.1\ $\Msun\ has been found within $2.6$ pc radius from
the cluster centre \citep{Luhman_2004}. The associated criterion is to have either zero or one of this 
kind of object left in the simulation.\\
The absence of very low-mass systems is observed for
both single objects and companions at a separation larger
than 50 AU. In our simulations, the number of VLMOs
is therefore the total number of companions (within a separation range of [50-400] AU),
single objects and close binaries (separation smaller than 50 AU) whose
mass is below 0.1 \Msun. 
\paragraph{Binary fraction ($N_5$)}
\citet{Brandeker_et_al_2006}
identified 5 binaries and 3 candidates for a total of 18 systems in
the core region. 
Considering the average value of 6.5 binaries, this
gives an observed binary fraction of 36\% and the validity range
for this criterion is from 22\% to 50\%.
Since binaries wider than 400 AU are
considered as two separate single stars in our analysis, the simulated binary
fraction is already of the order of $50$\% before any dynamical
evolution for models A, B, D and E
because of the initial period distribution.
Therefore this criterion is not expected to be critical.
\paragraph{Number of wide binaries ($N_6$)}
$\eta$ Cha does not contain any binary with a projected separation greater than 30 AU. This was put
into the following constraint : we require the model not to contain any binary with a separation larger 
than 50 AU. We choose a loose cut on separation to be more
conservative and to take projection effects into account. In
the following we refer to the number of wide binaries as the
number of binaries with separations larger than 50 AU and smaller than
400 AU.
\paragraph{Age}
With an initial state estimated to be between 0.1 to 3 Myr (see
section~\ref{Gas}), and an age for $\eta$ Cha taken to be 8 Myr, we
require the simulations to be in the age range from 5 to 8 Myr.
We also require the time window during which the other criteria
are fulfilled not to be smaller than 1 Myr, to exclude
transient states.
\subsection{Probability maps}
\label{proba_map}
Since it appears very difficult to satisfy all criteria simultaneously
for most models, we
refine our analysis and build a probability to estimate how likely a set of
simulations reproduces each observational constraint independently. At each time step
we compute the probability $a_i(t)$ for the simulation to fulfil
a criterion $i$.
This probability is calculated from the normalized
histogram generated from the simulations by summing all the bins in
the range $[O_i-\sqrt{O_i}, O_i+\sqrt{O_i}]$. Statistical scatter is
dealt with using a smoothed histogram in case of a poor bin
sampling. If none of the 200 simulations recover the range associated
to the observed value, we set the
probability to 1/200, regardless of the gap separating this interval
to the value of the first non-zero bin. In the case of a complete
mismatch between observation and model, this method does not provide
more information than an upper limit. \\
The probability $a_i(t)$ can
be calculated for each configuration \{$N_{\text{sys}}$,
$R_{\text{Pl}}$, $Q_i$\} and each model. In particular we can produce maps of
$a_i(t_{i,m})$ in coordinates of $N_{\text{sys}}$ and $R_{\text{Pl}}$
for a given $Q_i$ and a given model
(see e.g. Fig.~\ref{tend}), where $t_{i,m}$ corresponds to the time, in the
range [5,8] Myr, at which $a_i(t)$ is maximum.
\section{Results from our standard model (model A)}
\label{Res}
In this section, we discuss the results given by model A to test whether $\eta$
Cha can be reproduced from a universal log-normal {\it single} star IMF, with
100\% binary fraction and random pairing. This model is a first guess, based
on standard assumptions. The analysis presented below
motivated us to relax some assumptions (section \ref{Disc}).
\subsection{Reproducing $\eta$ Cha}
\label{QuantA}
\begin{figure*}
\centering
  \includegraphics[width=17cm]{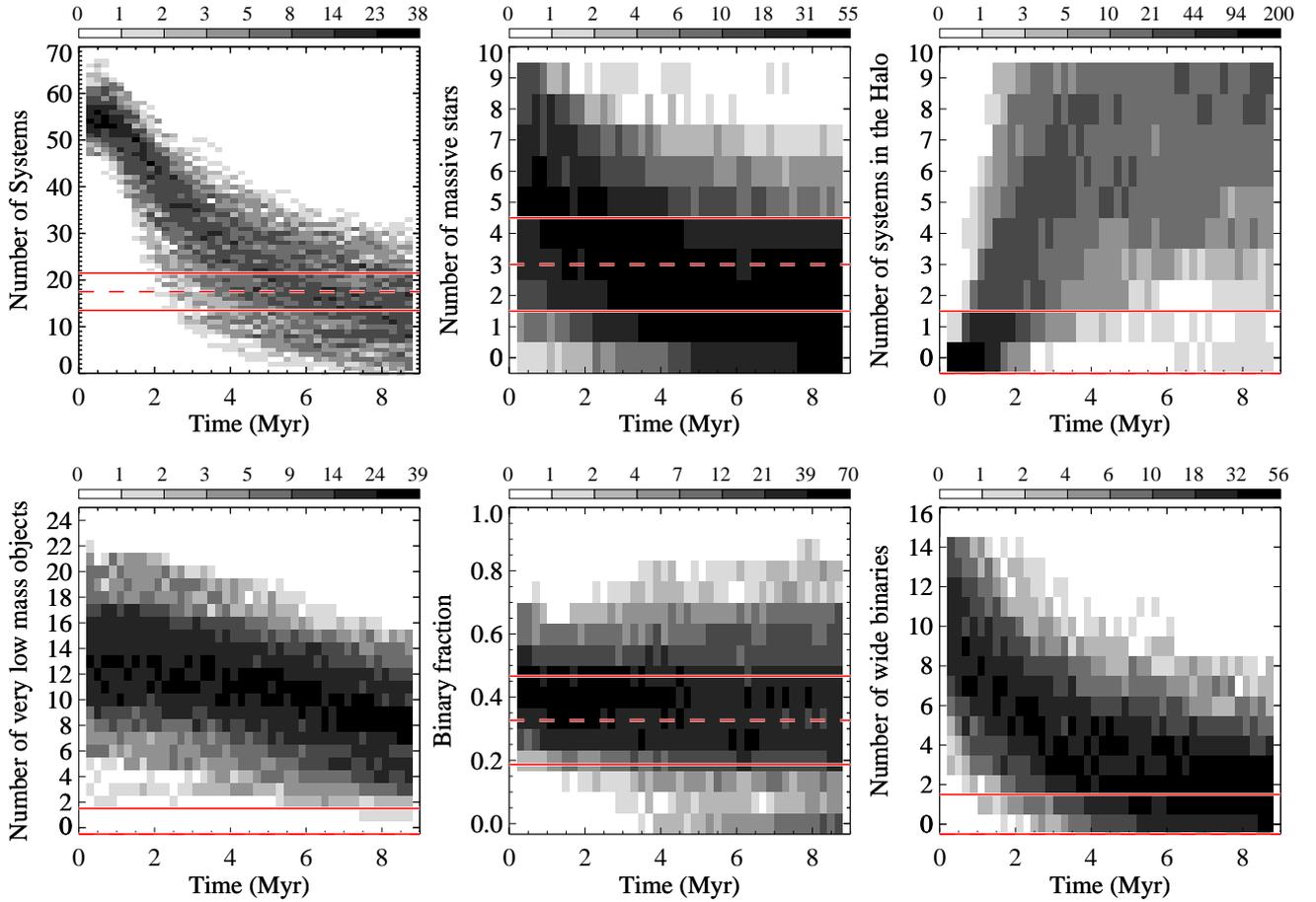}
  \caption{3D histograms showing the evolution of (a) the number of
    systems in the core $N_1$, (b) the number of massive stars in the
    core $N_2$, (c) the number of systems
    in the halo $N_3$, (d) the number of VLMOs inside a 2.6 pc radius $N_4$, (e) the binary
    fraction in the core $N_5$, and (f) the number of wide (separation greater than $50$
    AU) binaries $N_6$ as a function of time for model A and for
    the initial configuration \{$N_{\text{sys}}=40$,
    $R_{\text{Pl}}=0.05$ pc, $Q_i=0.5$\}. The histogram density
    corresponds to the number of simulations (out of 200) that fall in
    a given bin. The dashed and solid red
    lines correspond to the observed value and the acceptable range respectively.}
  \label{hist_all}
\end{figure*}

The criteria described in the previous section allow us, when used together, to check
the ability of model A to reproduce the observations for a given set
of initial parameters. Considering each of the 200 realizations for all
configurations \{$N_{\text{sys}}$,
$R_{\text{Pl}}$, $Q_i$\}, we apply these criteria at each time snapshot to see
if they can all be satisfied simultaneously. Table.~\ref{figstat} shows
a summary of this procedure for a specific value of virial ratio
($Q_i=0.5$) and number of systems ($N_{\text{sys}}=20$) for the first
4 criteria (thus without any constraint on the binary properties nor
the age). \\
\begin{table}
\caption{Results of the quantitative analysis for different values
     of $R_{\text{pl}}$, and for $N_{\text{sys}}=20$ and $Q_i=0.5$ for
     model A. For
     each value of $R_{\text{pl}}$ we apply the four first criteria one
     after the other. Every time a criterion is added, we compute the
     number of runs that fulfil the condition. As a result of our successive
     elimination scheme, only three simulations satisfies the first
     four criteria. However, none of those fulfil all six criteria simultaneously.}
   \label{figstat}
\centering
\begin{tabular}{l || c c c c c c}
\hline
Criterion & \multicolumn{6}{c}{$R_{\text{pl}}$} \\ 
 & 0.005 & 0.01 & 0.03 & 0.05 & 0.1 & 0.3\\ \hline \hline
Systems	& 195 & 199 & 197 & 199 & 199 & 200\\ \hline
+ Massive stars	& 96 & 114 & 139 & 150 & 154 & 143\\ \hline
+ Halo & 21 & 41 & 67 & 89 & 100 & 72\\ \hline
+ VLMOs & 0 & 1 & 1 & 0 & 1 & 0\\ \hline
\end{tabular}
\end{table}
Even if most runs satisfy the first
criterion on the number of systems, this is valid only for a given
time range, in which the next criterion will have to be fulfilled. The
most important result is that the percentage of runs passing the
selection drops to zero as we apply the fourth
condition on the number of VLMOs for all the initial configurations with 
$N_{\text{sys}}>20$, and to $0.5\%$ at best
for $N_{\text{sys}}=20$. \\
\begin{figure*}[ht!]
  \centering
  \includegraphics[width=17cm]{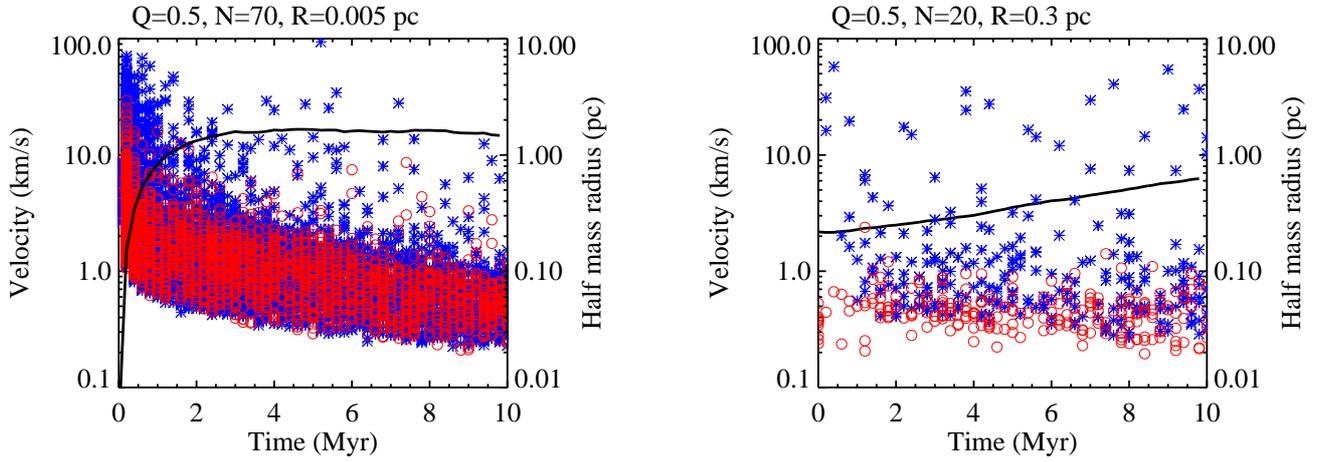}
  \caption{The left panel displays the velocity
    of ejected members at the time of ejection (blue asterisks for single
    objects, and red open circles for binaries) for a dense
    configuration from model A, with $N_{\text{sys}}=70$, $Q_i=0.5$,
    and $R_{\text{Pl}}=0.005$ pc. The right panel corresponds to a
    sparse configuration  with $N_{\text{sys}}=20$, $Q_i=0.5$,
    and $R_{\text{Pl}}=0.3$ pc). The half mass radius evolution
    is superimposed (thick line), as
    a footprint of the stellar density for both cases.}
  \label{VTall}
\end{figure*}
This indicates that this criterion is very difficult to fulfil 
simultaneously with the other criteria.
A dense initial state is necessary to remove all (or almost all) very low
mass members from the cluster by enhancing two-body encounters,
especially as more objects are released during the processing of wide
binaries. However this tends to quickly inflate the inner core, acting
in opposition to the criteria on the
number of systems ($N_1$) and massive stars ($N_2$) and
increasing the number of stars in the
halo ($N_3$). 
This is illustrated for the dense
initial configuration with \{$N_{\text{sys}}=70$,
$R_{\text{Pl}}=0.005$ pc, $Q_i=0.5$\} in the left panel Fig. ~\ref{VTall}.
There is a peak of cluster members ejection \footnote{
  ejected member are any object unbound to the cluster and being
  at a distance larger than twice the half mass radius from the
  cluster center} before 1 Myr with velocities as high as 60 km/s
(especially for single objects released by binary decay). As an
imprint of this highly dynamic phase the cluster undergoes a fast
expansion phase, shown by the increase of the half mass radius from
0.01 pc to 1 pc within 1 Myr. Once the density has fallen off, the
dynamics involves softer interactions (secular evolution) and the
number of ejected members decreases along with their velocity. During this phase
the cluster expands slowly until reaching virial equilibrium.\\
As a result of the fast relaxation phase the VLMOs are ejected efficiently
but the numbers of systems ($N_1$) and massive
stars ($N_2$) remaining in the cluster core are too small.
In addition, the core expansion adds many solar-type systems to the halo, incompatible with
the criterion $N_3$. We can move to a less
dense initial state to try to improve the results, but then the expansion
is too slow and the number of VLMOs inside a 2.6 pc radius ($N_4$)
remains almost constant with time. When starting with a sparse
configuration ($N_{\text{sys}}=20$,
$R_{\text{Pl}}=0.3$ pc and $Q_i=0.5$ ), we do not see any peak of ejection at earlier
times, and the half mass radius increases slowly and linearly in time
(Fig.~\ref{VTall}, right panel).
It seems therefore that a compromise on the initial density has to be
found in order to eject most of the VLMOs while retaining
a dense enough core (compatible with criteria $N_1$ and
 $N_2$) and without populating the halo.\\
To better understand the cluster dynamical evolution, we show in
Fig. ~\ref{hist_all} the evolution of the six quantities
constrained by the observations for the 200 realizations that started
with an intermediate density ($N_{\text{sys}}=40$, $Q_i=0.5$ and $R_{\text{Pl}}=0.05$ pc). The range
corresponding to each criterion is delimited by solid lines in each
panel.\\
First, it is interesting to note from the top left panel that the
number of systems does not actually start at the setup value 40, but around
53 in average. This is mainly due to not counting bound pairs with
separations larger than 400 AU as binaries but as two single objects,
thus increasing the total number of systems. This can also
be seen in the lower middle panel, where the binary fraction is
initially around $46\%$, instead of 100\% as set up.
Then, during the cluster early evolution phase, binaries are
processed more or less efficiently due to dynamical
interactions, depending on their separation and on the initial
density. In this case the binary fraction decreases from 46\%
to 43\% within 0.5 Myr.
As a consequence of the binary disruption the number of systems
inside the inner core increases slightly during the first
0.5 Myr. After this
phase, dynamical interactions are softer. The secular evolution
tends to inflate the core, slowly dispersing the cluster members,
decreasing the number of systems ($N_1$) and
wide binaries ($N_5$) in the inner core, and increasing the
number of stars in the halo ($N_3$). \\
The number of VLMOs inside a 2.6 pc radius ($N_4$) evolves
in a similar way to the number of systems in the inner core (see
bottom left panel). Note that the number of very low mass systems expected
from the initial conditions (log-normal \textit{single} star IMF, 100\% binary fraction
and random pairing) should be around
6 for $N_{\text{sys}}=40$. However, the number of VLMOs is already $\simeq
  13$ at $t=0$ Myr (bottom left panel of Fig.~\ref{hist_all}), due
to the fact that any very low mass ($m<0.1$ \Msun) companion at
separation larger than 400 AU is counted as a single object.  This
number remains constant during the first 0.5 Myr as binary
disruption compensates for the ejection process. Then, later in the cluster
evolution, the number of VLMOs ($N_4$) decreases slowly, as does the
number of systems. However, it remains larger than five for most
of the simulations starting with
$N_{\text{sys}}=40$, $R_{\text{Pl}}=0.05$ pc and $Q_i=0.5$.\\
Overall, the success rate for the simulations to reproduce the
observations is zero for all the initial configurations in model A.
\subsection{Best-fitting initial conditions}
\begin{figure}
  \resizebox{\hsize}{!}{\includegraphics{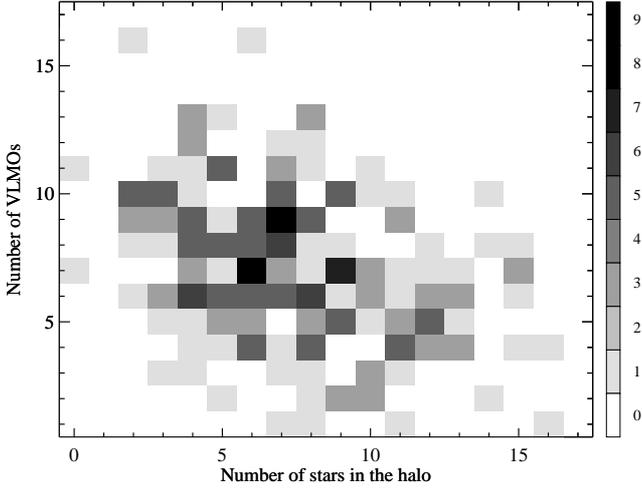}}
  \caption{Correlation map between the number of VLMOs ($N_3$)
    and the number of solar-type stars in the
    halo ($N_4$) at $t=7$ Myr for
    $N_{\text{sys}}=30$, $R_{\text{Pl}}=0.05$ pc and $Q_i=0.5$ (model
    A).}
  \label{correl2}
\end{figure}

\begin{figure}
  \resizebox{\hsize}{!}{\includegraphics{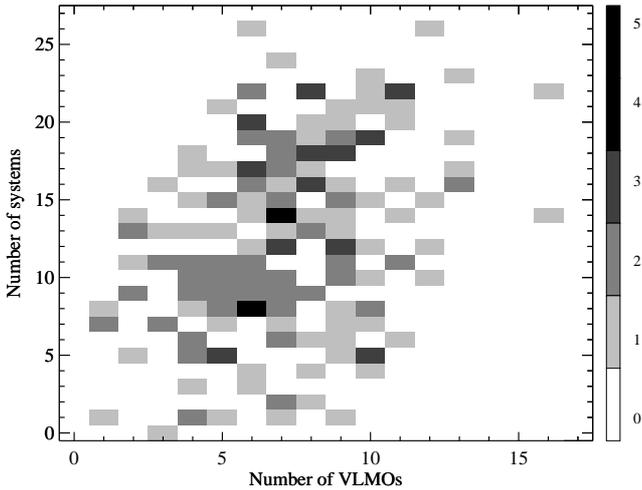}}
  \caption{Correlation map between the number of systems ($N_1$) and the
    number of VLMOs ($N_3$) at $t=7$ Myr for
    $N_{\text{sys}}=30$, $R_{\text{Pl}}=0.05$ pc and $Q_i=0.5$ (model
    A).}
  \label{correl}
\end{figure}

In order to know which criterion is the most stringent and how the
model hypothesis could be modified to
reproduce $\eta$ Cha, we performed the analysis based on probability maps of
$a_i(t_{i,m})$ described in section \ref{proba_map}.
\begin{figure*}
\centering
  \includegraphics[width=17cm]{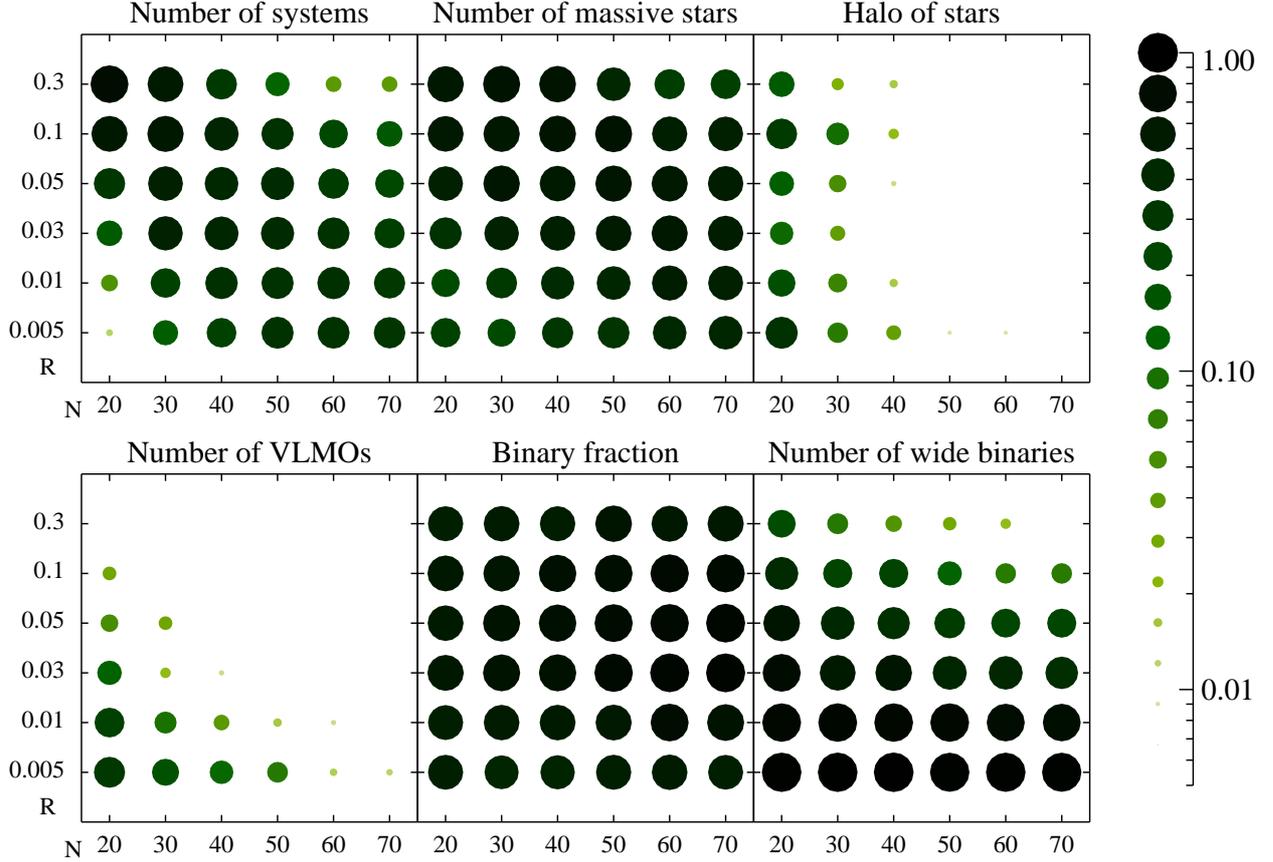}
  \caption{Summary for all tested configurations for model A with a virial ratio
    $Q_i=0.5$. For convenience the value for the quality measure,
    $a_i(t_{i,m})$ is also indicated by the dot size, with a logarithmic
    scale. For each criterion, the colour (and size) of the circles
    indicates the probability for the simulations to reproduce the
    observations. The results obtained with a different virial ratio
   $Q_i $ are very similar.}
  \label{tend}
\end{figure*}
Fig.~\ref{tend} reveals the regions (in \{$N_{\text{sys}}$,
$R_{\text{Pl}}$\} coordinates, for $Q_i=0.5$) which most likely satisfy a given
observational criterion. Below we review the inability of
the simulations performed in model A to reproduce the observations,
in the light of Fig.~\ref{tend}. Note that the time constraint is
not mentioned, but was applied to produce all the  discussed probability maps.
\paragraph{$N_1$}
We notice that the probability to fulfil the criterion
on the number of systems in the core drops if we
start with large initial values for
$N_{\text{sys}}$ and $R_{\text{Pl}}$ since the density becomes too
small to remove enough systems from the inner core by dynamical
interactions. On the other hand, when starting with a low
$N_{\text{sys}}$ and $R_{\text{Pl}}$, the number of systems that
remain in the core becomes
rapidly too small.\\
\paragraph{$N_2$}
The criterion on the number of massive stars seems to be easy to
reproduce and does not strongly depend on the initial parameters
although there is a small trend in favour
of less dense cases or large value of $N_{\text{sys}}$.\\
\paragraph{$N_3$}
Considering the number of systems in the halo, it is clear that
this criterion is best matched with the smallest $N_{\text{sys}}$ because less
objects can be ejected in the halo. For $N_{\text{sys}}=30$ or 40,
this criterion is more easily fulfilled for either
large $R_{\text{Pl}}$ (as lower density leads to fewer ejections), or
small $R_{\text{Pl}}$ (as high density induces fast ejections, leading
to large projected distances by 5 Myr). Intermediate values of
$R_{\text{Pl}}$ result in too many slow-moving ejected stars that will
remain in the vicinity of the cluster. For
$N_{\text{sys}}>40$ the criterion on ($N_3$) is very badly
reproduced for any value of $R_{\text{Pl}}$.\\
\paragraph{$N_4$}
The result for the VLMOs is important since the region of
agreement is very narrow. This shows that
this criterion together with $N_3$, is the most stringent. It requires
a low value of $N_{\text{sys}}$ to minimize the initial number of VLMOs
to eject, and a low $R_{\text{Pl}}$ to maximize the
dynamical encounters and eject these objects efficiently.\\
\paragraph{$N_5$ and $N_6$}
We notice that the map for the binary fraction ($N_5$) does not indicate
a large dependence on the parameters with an overall good agreement with
the observations. The separation map ($N_6$) reveals a higher probability
for the dense cases, which process the widest binaries and expand
fast enough so that less binaries are present in the central region.\\
\paragraph{} 
Although very narrow, the overlap region between the agreement maps of the various
criteria (especially those for $N_3$ and $N_4$) seems 
to indicate that suitable initial
configurations may be found (e.g see Fig. \ref{tend})
for intermediate to low $N_{\text{sys}}$ and low
$R_{\text{Pl}}$ (except for the lowest values for which the criterion
on the number of systems $N_1$ is not well fulfilled). 
However this result is misleading for the
criteria are not independent. 
There is a significant anti-correlation
between the number of stars in a 10 pc halo ($N_3$) and the number of VLMOs
in a 2.6 pc radius ($N_4$) at low and 
medium initial densities (as seen Fig.~\ref{correl2} for
$N_{\text{sys}}=30$ and $R_{\text{Pl}}=0.05$ pc).  In these cases the constraints on
$N_3$ and $N_4$ tend not to be compatible. At higher densities and
especially for $R_{\text{Pl}}=0.005$ pc, this anti-correlation becomes
negligible. Both $N_3$ and $N_4$ get very small: the strong
dynamical interactions remove all VLMOs from the core, and most
ejected objects travel much further away than 10 pc within 5 Myr due
to the high ejection velocities. However, the dynamical interactions
are so strong that it is very difficult to retain anything in the cluster core and the
number of systems $N_1$ becomes too small. Fig.~\ref{correl} shows the correlation
between $N_1$ and $N_4$ at
$t=7$ Myr for the 200 simulations starting with $N_{\text{sys}}=30$
and $R_{\text{Pl}}=0.05$ pc. In all cases, when the number of VLMOs is
less or equal to one, the number of systems in the core is smaller than
10, making these two criteria incompatible. We found a similar correlation
for all the initial configurations tested by our
simulations. 
The negative result of the previous analysis indicates that ejecting all
VLMOs from the cluster core and keeping enough
systems in a $0.5$ pc sphere is a major
challenge. \\
The results obtained for different values of $Q_i $ (from 0.3 to
  0.7) cannot be statistically distinguished from those obtained for
  the state initially in virial equilibrium ($Q_i=0.5$).
\subsection{Summary and comparison with the previous study}
Using standard initial conditions corresponding to model A, we tested many different
configurations, varying the density and virial ratio. The
main conclusions from this analysis are that
\begin{itemize}
\item starting with a \textit{single} log-normal IMF with a peak value
  $\mu = 0.2$ \Msun\ and a deviation $\sigma = 0.55$, and assuming an
  initial binary fraction of 100\%, random pairing and a Kroupa-like
  period distribution for the binary population, does not allow the
  simulations to reproduce the observations for any configuration
  \{$N_{\text{sys}}$, $R_{\text{Pl}}$, $Q_i$\};
\item there is no hint of an improvement at the edge of the
  parameter grid, suggesting that our failure to find a solution
  is not a consequence of using a limited parameter space.
\end{itemize}
In \citet{Moraux_Lawson_Clarke_2007} the best fitting set of initial
parameters gave a success rate of about 5\%, whereas in our analysis of
model A it is 0\%. The apparent divergence between our results and
Moraux's is the consequence of the initial conditions. In the previous
study, the chosen IMF corresponded to the {\it system} mass function
obtained after binary processing (as it is observed in the field or in
clusters), and binaries were considered as unbreakable objects, unable
to exchange energy to the cluster by modifying their orbital
properties. Here, the system IMF peaks at higher masses, generating
more systems with mass $m>0.5$ \Msun ~initially, potentially increasing
the number of them that could end up in the halo. This makes
the criterion on $N_3$ more difficult to fulfil in the present
study.
Besides, binary disruption can significantly alter our ability
to reproduce criterion $N_4$. 
Even though there are less very low mass {\it systems}
initially, many objects with $m<0.1$ \Msun\ ~belong to a binary system
with a separation larger than 50 AU or have been be released by binary
decay. In both cases, these objects will be accounted for in the
number of VLMOs ($N_4$), and this criterion is therefore not improved.
\section{Alternative initial conditions}
\label{Disc}
We discussed above the importance of the binary population in shaping
the system IMF and hosting
VLMOs that may be released in the cluster core. Since these processes depend strongly
on the binary properties (mass ratio and separation distributions), we
will now describe how they may be adjusted (model B and C) to better reproduce the
observations. We will also discuss the possibility that the \textit{single}
star IMF might be discontinuous around the substellar limit (model
D), which may help to reduce the initial number of VLMOs in the
core. We will then present the results obtained in the extreme
case when starting with an IMF truncated at 0.1 \Msun\ and a binary
fraction of 100\% (model E) or less (model F).
\subsection{Binary pairing (Model B)}
\label{modelB}
In model A, we chose for simplicity to pair binaries randomly from the same
single IMF. Nevertheless, recent studies of both
the galactic field \citep{Raghavan_et_al_2010, Reggiani_Meyer_2011} and star forming regions
\citep{Kraus_et_al_2008, Kraus_et_al_2011} indicate that, whereas there is no
clear and unique best fit, a flat mass ratio distribution may be a better fit than a
random pairing.\\
Since this would result in a slightly smaller number of very low mass
companions, we may expect the criterion on $N_4$ to be better fulfilled. To implement it in our initial
conditions for model B, we sample the primary mass from a {\it primary} IMF and
then draw the secondary mass according to a flat mass ratio
distribution. This requires a slight change in the parameters
of the \textit{primary} IMF, in order to reproduce
the log-normal {\it single} IMF. This gives $\mu=0.32$
\Msun\ and $\sigma=0.55$ (instead of $0.2$ \Msun\ and $0.55$ in case of
random pairing). The binary fraction and
separation distribution are the same as in model A.\\
We ran simulations for $Q_i=0.5$ only,
$N_{\text{sys}}$=20 and $N_{\text{sys}}$=40, with the same range for
$R_{Pl}$ as before.
Results show that the agreement probability on the number of VLMOs
($a_4$) is larger (by a factor of two to three),
contrary to the probability $a_3$ on the halo that is smaller (by a factor of
about 2). This is due to the shift towards higher masses of the
primary MF yielding more objects with $m>0.5$ \Msun ~that may end up in
the halo. Overall, no significant improvement
is observed when using a flat mass ratio distribution since again
no simulation is able to reproduce the observations.\\
\subsection{Separation distribution (Model C)}
\label{SepDist}
\begin{figure}
  \resizebox{\hsize}{!}{\includegraphics{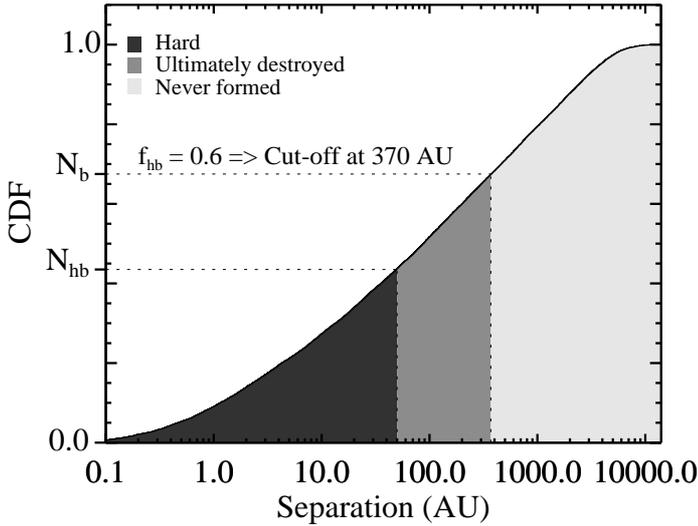}}
  \caption{Cumulative separation distribution function (solid line) obtained from
  the Kroupa period distribution (see section \ref{In}). This highlights
  how the distribution is truncated in Model C in the case of $f_b=0.8$. Expression
  \ref{binfraction} yields $f_{hb}=0.6$, which imposes to truncate the separation
  distribution to 370 AU, so that $N_{hb}=0.6 \times N_b$.}
  \label{sepini}
\end{figure}

\begin{figure}
  \resizebox{\hsize}{!}{\includegraphics{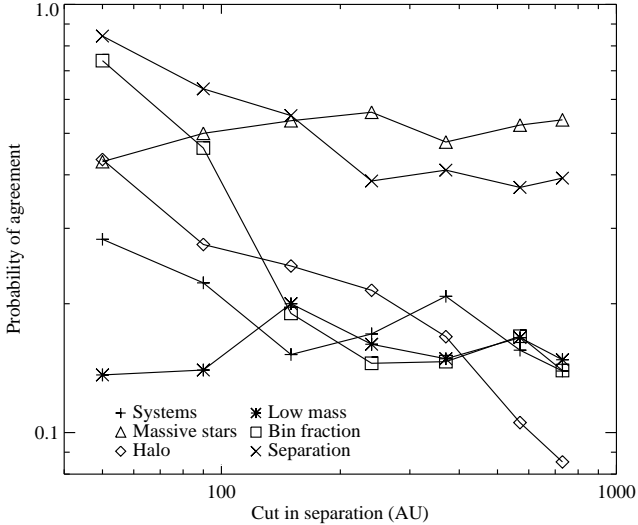}}
  \caption{Evolution of the agreement probability for the different
    criteria, when setting a maximum value for the binary separation,
    varying from 50 to 730 AU, and adopting a flat
    mass-ratio distribution (model C). The results are given for \{$Q_i$=0.5,
    $N_{\text{sys}}$\ =\ 20, $R_{pl}$\ =\ 0.1 pc\}.  }
  \label{A_evol}
\end{figure}

In the following, we discuss the possibility that no
wide binary formed initially by assuming an initial
separation distribution similar to model A but truncated at large separation. The cut-off separation value and the initial binary
fraction are linked to each other and we explain below how they can be evaluated providing the final binary fraction. \\
Following the simplistic argument that all binaries with a separation
smaller than a given value (hard binaries) survive
throughout the simulation, and that any wider binary is destroyed, we
can express the initial binary fraction $f_b$ in terms
of the initial hard binary fraction $f_{hb}=N_{hb}/N_{b}$ (where $N_b$
is the number of binaries and $N_{hb}$ the number of hard binaries) and the
final binary fraction $f'_b$: 
\begin{equation}
\label{binfraction}
  f_b=\frac{f'_b}{(1+f'_b)f_{hb}-f'_b}
\end{equation}
Taking a final binary fraction of 36\% given by the observations,
we consider different values
of $f_b$, ranging from 36\% to 100\%. The corresponding initial hard
binary fraction ranges from 100\% (all binaries survive) to 53\%
(about half the binaries survive) respectively. To follow the observations, we identify as
hard binaries (that will not be destroyed) those with separations lower than 50 AU. From the
initial hard binary fraction, we then estimate the corresponding
separation cut-off, assuming a \citet{Kroupa_1995c} distribution
below this value.
For example, we need a cut-off at 730 AU for a hard binary
fraction of 53\%. The lowest possible cut corresponds naturally to 50 AU, to
get 100\% hard binaries. Fig.~\ref{sepini} illustrates this process
in the case of an initial binary fraction of $f_b=0.8$, which gives $f_{hb}=0.6$.
This initial hard binary fraction is obtained when applying a cut-off in the separation
distribution around 370 AU.\\
\begin{table}
\caption{Binary fraction and separation cut-off for model C.}
\label{IMFtable}
\centering
\begin{tabular}{c | c c c c c c c}
\hline
    $f_b$ (\%) & 100 & 90 & 80 & 70 & 60 & 50 & 40 \\
    \hline
    Cut-off (AU) & 730 & 570 & 370 & 240 & 150 & 90 & 50 \\
    \hline
\end{tabular}
\end{table}
We can wonder why these binary properties would result from the cluster
formation process and this needs to be compared to what is observed in
star forming regions. In dense environment such as the Trapezium
$f_b\simeq$ 60\%, whereas in sparse regions like Taurus $f_b\simeq$ 90\%
\citep{Duchene_1999, Kirk_Myers_2012}. A
plausible explanation for this difference is that all star forming regions start their
evolution with a high binary fraction and the wide binaries are
further disrupted in dense environments within 1 Myr \citep[see e.g][]{Marks_Kroupa_2012}.
In the Trapezium, very
few binaries with a separation larger than 1000 AU have been found
\citep{Scally_Clarke_McCaughrean_1999} which supports this
idea. For instance it was possible to reproduce the evolution of the ONC
\citep{Kroupa_Aarseth_Hurley_2001,Marks_Kroupa_2012} starting 
with 100\% binaries and a density of $10^5$ stars/pc$^3$, including the
deficit of [200-500] AU binaries compared to
the separation distribution for field binaries from \citet{Raghavan_et_al_2010} \citep{Reipurth_et_al_2007}.
In our simulations starting with $N_{\text{sys}}=20$ and
$R_{Pl}=0.05$ pc, the initial density is very similar
($2 \times 10^5$ stars/pc$^3$). However, the adopted separation cut-off at 50
or 90 AU cannot be explained by dynamical encounters since these
separation limits are much lower than the initial mean neighbour 
distance (around 2200 AU).
Nevertheless, it is still possible that the binary
fraction may be set up during the formation process and/or
during the gas-rich phase which is not covered in our simulations.\\
We ran the simulations for $Q_i=0.5$, $N_{\text{sys}}= 20$ to 70, and
$R_{Pl}=0.005$ pc to 0.3 pc (model C). The parameters used for the
binary fraction and separation cut-off are given in
Table~\ref{IMFtable}. A flat mass ratio distribution (as in model B)
has been used to generate the secondary masses.\\
Fig.~\ref{A_evol} shows the evolution of the probability
$a_i(t_{i,m})$ in the case $N_{\text{sys}}=20$ and $R_{Pl}=0.1$ pc for
the six criteria as a function of the adopted separation
cut-off. For a large separation cut-off, the probability of
  agreement for the binary fraction is low ($<0.2$). This is worse
  than what was obtained for models A and B, for which no cut-off was
  applied to the Kroupa-like separation distribution. This is because
  (1) more binaries have a separation lower than $400$
  AU and are thus identified as binaries in the analysis procedure
  leading to a higher initial binary fraction, and (2) the high binary
  fraction remains almost constant in time, unless the initial
  density is very high.  An improvement is naturally seen for the
  criteria on the binary fraction as well as on the number of wide
  binaries when the separation limit gets smaller ($<100$ AU).
  Applying a cut-off at 50 AU corresponds to removing the constraints
  on the binary population since we already start with what is
  observed (no wide binaries and $f_b=40$\%).  The probability
  $a_3$ for the halo is also increasing, from 0.08 to
  0.4 for the lowest cut-off. For the number of
  systems ($N_1$), the number of massive stars ($N_2$), and the number of
  VLMOs ($N_4$) the probability does not change significantly. This may
  be surprising at first, especially for $N_4$, as
  less VLMOs will be produced by binary decay. However, this effect is
  compensated by the slower dynamics making it more difficult to eject
  the VLMOs from the core even though they are less numerous. \\
Nevertheless, the analysis reveals two configurations ($Q_i=0.5$, $N_{\text{sys}}=20$,
and $R_{Pl}=0.05$ pc and 0.1 pc)
for which some simulations satisfy all criteria
if the separation cut-off is 50 AU. 
We found respectively one and three simulations out of 200 that fulfil
all the observational constraints.\\
To check whether
these successful runs are consistent with the recent results from
\citet{Murphy_Lawson_Bessell_2010}, we look at the number of low mass
systems in the mass range [0.08, 0.3]~\Msun\ located at a distance range [2.6, 10]~pc from the
cluster centre. We find between zero to one of these systems, which is possibly too small
compared to the detection of four probable
plus three possible candidates.
\subsection{Treating brown dwarfs as a separate population (Model D)}
So far we have considered a continuous IMF that extends to the substellar
regime (down to $0.01$ \Msun), but 
\citet{Thies_Kroupa_2007} suggest that brown dwarf (BD) formation may be different to
star formation (based on their binary properties), which would lead to a
discontinuous mass distribution for single objects.
This assertion is still a matter of debate, but nonetheless
finds observational support\footnote{A recent review \citep{Jeffries_2012}
emphasize that the lack of coherence and completeness of the observations
do not allow firm conclusions.}
from the mass function of young star
clusters \citep{Thies_Kroupa_2008} and BD binaries surveys
\citep{Kraus_Hillenbrand_2012}. \citet{Parker_Goodwin_2011} excluded
pure dynamical evolution as a possible explanation for the observed
differences between the separation distributions of stellar and
substellar binaries, implying that it may be a pristine feature (or set during the
very early evolution).
From a theoretical point of view, the
process of BD formation remains unclear and may involve a star-like
collapse within a turbulent medium \citep[e.g.][]{Whitworth_Stamatellos_2006}
or a more specific channel of early ejection of gaseous clumps
\citep{Reipurth_Clarke_2001,Basu_Vorobyov_2012}. Other plausible mechanisms suggest massive disc
fragmentation \citep{Stamatellos_et_al_2007} or gravitational instabilities
induced in disks as a result of encounters in embedded clusters \citep{Thies_et_al_2010}.\\
To evaluate the possibility that the {\it single} IMF may be discontinuous
(model D), we consider
initial conditions corresponding to the results from \citet{Thies_Kroupa_2008}.
We adopted two log-normal single mass functions with $\mu=0.08$ \Msun~
and $\sigma=0.69$, but one corresponds to stars and is limited to the mass
range [0.07, 4]~\Msun\ and the other one corresponds to BDs and very low mass stars (VLMS) with
$0.01<m<0.15$ \Msun. There is an overlap between the
two mass functions in order to end up with a continuous {\it system}
IMF consistent with the universal picture of the IMF. Each population (stars, and BDs + VLMSs) is treated
separately and the BDs and VLMSs to stars ratio is assumed to be
$1/5$. The BD and VLMS binary fraction and the star binary fraction are respectively 30\% and
100\% and there are no mixed BD/VMLs binaries \citep{Kroupa_et_al_2011}.
For simplicity we generate the binaries for
each population using random pairing and the same period
distribution with no separation cut-off. The latter hypothesis is
not realistic since field BD binaries are known to have a tighter period distribution \citep{Burgasser_et_al_2007}
that cannot be explained by pure dynamical evolution \citep{Parker_Goodwin_2011}.
Nevertheless this will have a very limited impact on our results, since the number
of BD binaries is one or two in average (if starting respectively with $N_{\text{sys}}=20$
or $N_{\text{sys}}=40$). \\
We ran simulations in the virialized case for $N_{\text{sys}}=20$ 
and $R_{\text{Pl}}$ within \{0.05, 0.1\} pc as well as for $N_{\text{sys}}=40$
and $R_{\text{Pl}} = 0.05$ pc.
As a result of the analysis the improvement over our previous
simulations is limited: no simulation matches all observations of the
$\eta$ Cha cluster. Compared to the standard case 
(model A), the main improvement lies in the probability $a_3$ to
reproduce the halo, which is mainly explained by the shift towards
  lower masses of the system IMF. At best the probability increases from 0.29 to 0.55
in the case with $N_{\text{sys}}=20$ and $R_{\text{Pl}} = 0.1$ pc.
Note that this probability is also higher than
when we applied a separation cut-off at 50 AU (model C, see Fig.\ref{A_evol}). 
However, the probability of agreement for the criterion on the
VLMOs decreased compared to the result from model A (from 0.03 to
0.005 for $N_{\text{sys}}=20$ and $R_{\text{Pl}} = 0.1$ pc).
We can understand this by counting the mean
number of VLMOs at t=$0.6$ Myr, after the binary breaking phase: for
$N_{\text{sys}}=20$ and $R_{\text{Pl}} = 0.1$ pc we find $N_{VLMO}=8$,
compared to $N_{VLMO}=6.5$ in the standard
case (model A) and to $N_{VLMO}=4.2$ in the extreme case of model C starting with
$40$\% binaries and a cut-off at $50$ AU in separation.
Since the number of VLMOs is a strong constraint, this comparison shows the
limited effect of the changes adopted for the BD population.
\subsection{Truncated IMF at the low mass limit and truncated separation distribution (Models E and F)}
The previous analysis indicates that the observational result regarding the number of VLMOs in $\eta$ Cha
is particularly difficult to reconcile with the other constraints, in
particular with the number of systems in the core and the absence of
solar-type star in the halo (see Fig.~\ref{correl2} and~\ref{correl}).
In the following
we consider the extreme scenario where the IMF is {\it
not} universal and no very low mass system
($m<0.1$\ \Msun) has formed initially. To do so, we generate primary
masses from  the same {\it primary} IMF (peaking
at $0.3$\ \Msun) as in model B, but truncated at $0.1$ \Msun , and use
a flat mass ratio distribution (without any truncation on the secondary
mass)\\
We first ran simulations
starting with a virialized Plummer sphere with 100\% binaries drawn from the Kroupa
separation distribution (model E). $N_{\text{sys}}$ and $R_{\text{Pl}}$ are chosen within
\{20, 30, 40, 50, 60\} and 
\{0.005, 0.01, 0.03, 0.05, 0.1, 0.3\} pc respectively. We discarded the larger value
$N_{\text{sys}}=70$ since it would give a cluster starting with too many systems to fulfil the criteria on
the number of systems in the inner core without populating the
halo. As a result no simulation fulfilled all 6
criteria. This as an outcome of both the truncation itself and the choice
for the initial binary fraction of 100\%. Because of the lower mass limit
the initial number of stars with $m>0.5$ \Msun\ increases for a given $N_{\text{sys}}$, making the criterion
on the halo more difficult to reproduce. In addition, since there is no truncation
in the secondary mass distribution, a few VLMOs are part of a
binary system and will appear as single objects, either because their
separation is larger than 400 AU or because the binary will be
processed by dynamical evolution. For instance, in the case $N_{\text{sys}}=20$
and $R_{\text{Pl}}=0.1$ pc,
$2.5$ VLMOs are identified in average at $t=0$ Myr. As a consequence the
criteria on the VLMOs and on the halo remain difficult to fulfil together with the other
criteria.\\
We ran additional simulations (model F) where we introduced a cut-off in the initial binary
period distribution in a similar way as for model C (see section \ref{SepDist}). We find that when starting
with a binary fraction of 40\% and a cut-off at 50 AU many more simulations could
reproduce the six observational constraints with a success rate up to 10\% for $N_{\text{sys}}=20$
and $R_{\text{Pl}}=0.1$ pc. The average number of VLMOs in this initial configuration is $0.2$ at $t=0$ Myr,
which shows that the constraint on the number of VLMOs is easily satisfied, given that there are only close binaries
(with separation $<50$ AU) that very stable dynamically.
 It is interesting to note that the very
 few runs of model C that are also able to
 reproduce all the criteria correspond to the same initial
 conditions. Indeed the successful cases were obtained for $N_{\text{sys}}=20$
 and $R_{\text{Pl}}=0.05$ or 0.1 pc and started with only one or two VLMOs
 initially.
This seems to indicate that both the IMF {\it and} the initial binary
population of $\eta$ Cha were not standard.\\
When considering Murphy's constraint however, the success rate shrinks
to 0.5\% at best, if we require to have at least three stars in the mass range [0.08; 0.3] \Msun\ 
and within a 10 pc radius. Despite a low success rate, this model
is the only one that can reproduce all the observational
constraints, including Murphy's results. Note that the only successful runs are for two medium density
configurations: \{$N_{\text{sys}}=30$; $R_{\text{Pl}}=0.1$ pc\} and
\{$N_{\text{sys}}=30$; $R_{\text{Pl}}=0.05$ pc\}.
\section{Summary and conclusion}
We have conducted a large set of pure N body simulations that
aim to reproduce the peculiar properties of the $\eta$ Cha
association, namely the lack of very low mass objects ($m<0.1$ \Msun)
and the absence of wide binaries (with a separation $>50$ AU).
We tested several models of various IMF and binary
properties, and span the parameter space in density and virial ratio.
The analysis was done using
several procedures in order to compare
efficiently the simulation results with the observational data and
identify the best initial state.\\
In order to test a universal picture for the IMF, we
assumed a continuous log-normal {\it single} IMF with $\mu=0.2$ \Msun
and $\sigma=0.55$.
Starting with this IMF and a binary fraction of 100\% (with either a
random pairing, model A, or a flat mass ratio distribution, model B), the
analysis shows that ejecting all very low mass members without creating
a halo of solar-type stars and keeping an inner core of 18 systems
is not possible. Similarly to the case of a discontinuous {\it single} IMF,
no simulation was able to match the observations.\\
Reproducing all available observations of $\eta$ Cha by pure
dynamical evolution from an universal {\it single} IMF {\it and} a stellar
binary fraction of 100\% is therefore very unlikely.\\
We then tested a different set-up for the binary
population, while preserving the shape of the IMF, our working
hypothesis (model C). We assumed that wide binaries do not form
initially and
adopted a separation distribution truncated at large separation
resulting in a lower binary fraction.
As a result, the best initial state, starting with an initial binary fraction of
40\% binaries and without any binary wider than 50 AU, yields a small success
rate of 1\% (that drops to 0\% if we require 
those simulations to have a halo of ejected low mass stars
\citep{Murphy_Lawson_Bessell_2010}).
Since almost no considered initial state assuming a universal IMF
statistically matches the 
observational constraints, we started with a truncated IMF
with no system below
0.1 \Msun. However, this fails in reproducing the observations, unless
starting with a singular binary population (no wide binary and a small binary
fraction; model F).
In this case, the best success rate is 10\% and is obtained for
initial parameters ($N_{\text{sys}}=20$
and $R_{\text{Pl}}={0.1; 0.05}$ pc) that are very similar to what is observed today in
the cluster.\\
This suggests that the dynamical
evolution did not play a strong role in shaping the properties of
$\eta$ Cha and that most of them must be pristine. $\eta$ Cha may have
started with an IMF deficient in VLMOs {\it and} with peculiar binary
properties (namely
a small binary fraction and an orbital period distribution
truncated at small periods). Note that
this conclusion is very different from \citet{Moraux_Lawson_Clarke_2007}
where the initial high density case was the preferred solution. This stresses
the importance of the binary population in the overall dynamical evolution of
the cluster.\\
One can speculate onto the particular physical conditions that might have produced
so few VLMOs together with preventing wide binaries from forming.
$\eta$ Cha may for instance originate from a highly magnetized cloud, preventing fragmentation
of large scale \citep{Hennebelle_et_al_2011}, forcing more mass into single fragments
and not creating wide systems. Tighter binaries could then be produced later on, after
the magnetic field has diffused out.\\
Finally, in the low density case solution presented above,
it is very difficult to reproduce the
recent results from \citet{Murphy_Lawson_Bessell_2010}. When
considering this additional constraint, the success rate becomes very
small (0.5\% at best).
Additional knowledge of the kinematics of this purported halo population might help
refine the dynamical picture of $\eta$ Chamaeleontis.
\begin{acknowledgements}
The authors wish to thank S.~Aarseth for allowing us access to his
N-body codes. We also thank A. Bonsor for her help to improve the manuscript,
and C.~Clarke, S.~Goodwin for useful
discussion and comments. This research has been done in the
framework of the ANR 2010 JCJC 0501-1``DESC''. The computation presented in
this work were conducted at the \textit{Service Commun de Calcul Intensif de 
l'Observatoire de Grenoble} (SCCI), supported by the ANR contract 
ANR-07-BLAN-0221, 2010 JCJC 0504-1 and 2010 JCJC 0501-1.
\end{acknowledgements}
\bibliographystyle{aa}
\bibliography{all_bibli}
\end{document}